\documentclass[12pt]{article}
\usepackage{amsmath,amssymb,graphicx,mathrsfs,hyperref,slashed}
\newcommand{\be}{\begin{equation}}
\newcommand{\ee}{\end{equation}}
\newcommand{\bea}{\begin{eqnarray}}
\newcommand{\eea}{\end{eqnarray}}

\def\({\left(} \def\){\right)}

\renewcommand{\baselinestretch}{1.5}
\begin{document}
\title{\vspace{-1.8in}
{Quantum state of the black hole interior}}
\author{\large Ram Brustein${}^{(1)}$,  A.J.M. Medved${}^{(2,3)}$
\\
\vspace{-.5in} \hspace{-1.5in} \vbox{
 \begin{flushleft}
  $^{\textrm{\normalsize
(1)\ Department of Physics, Ben-Gurion University,
    Beer-Sheva 84105, Israel}}$
$^{\textrm{\normalsize (2)\ Department of Physics \& Electronics, Rhodes University,
  Grahamstown 6140, South Africa}}$
$^{\textrm{\normalsize (3)\ National Institute for Theoretical Physics (NITheP), Western Cape 7602,
South Africa}}$
\\ \small \hspace{1.07in}
    ramyb@bgu.ac.il,\  j.medved@ru.ac.za
\end{flushleft}
}}
\date{}
\maketitle
\begin{abstract}

If a black hole (BH)  is initially in an approximately pure state and it evaporates by a unitary process, then the  emitted radiation will be in a highly quantum state.  As the purifier of this radiation, the state of the BH interior must  also be in some highly quantum state.  So that, within the interior region, the mean-field approximation cannot be valid and the state of the BH cannot be described by some semiclassical metric.  On this basis, we  model the state of the BH  interior as a collection of a large number of excitations that are packed into closely spaced but single-occupancy energy levels; a sort-of ``Fermi sea'' of all light-enough particles. This highly quantum state is surrounded by a  semiclassical region that lies close to the horizon and has a non-vanishing energy density. It is shown that such a state looks like a BH from the outside and decays via gravitational pair production in the near-horizon region at a rate that agrees with the Hawking rate. We also consider the fate of  a classical object that has passed through to the BH interior and show that, once it has crossed over the near-horizon threshold, the object meets  its demise extremely fast. This result cannot be attributed to a  ``firewall'', as the trauma to the in-falling object only begins after it has passed through the near-horizon region and enters a region where semiclassical spacetime ends but the energy density is still parametrically smaller than Planckian.

\end{abstract}
\newpage
\renewcommand{\baselinestretch}{1.5}\normalsize

\section{Introduction}

What does the interior of a black hole (BH) look like?
Leaving aside the wildly speculative, one  is left with three options: (1) Most of the matter is piled up near the center of the BH, as might be expected if the classical general-relativistic picture is approximately correct, (2) the matter is  concentrated  in the proximity of  the horizon in a highly excited state, as would be argued by the  proponents of ``firewalls'' and  ``energetic curtains''
 \cite{AMPS,Braun,Sunny,MP,Bousso} as well as ``fuzzballs''
\cite{Mathur1,Mathur2,Mathur3,Mathur4,Mathur5}
  and (3) the matter is distributed (more or less) uniformly throughout the
interior region. These options are depicted in Fig.~1.

The first of these options can be  promptly dismissed on the grounds that the state of the BH can be highly degenerate, as follows from  the large BH entropy $S_{BH}$ \cite{Bek}. This amount of information  cannot be stored in a region of spacetime that is any smaller than that bounded by the BH horizon itself, meaning that the first choice is really a remnant in disguise.  Another argument against the first option is that such a configuration would be highly unstable due to pair production occurring  deep in the BH interior, where the gravitational field would be strongest ({\em cf}, Section~3). The situation then boils down to  deciding between the second and the third options: Is there a firewall or a fuzzball surface at the horizon or isn't there? If there is a firewall or a fuzzball surface, then the inside, for all practical purposes,
no longer exists. If there isn't, then what is the state of the interior?

\begin{figure}
[t] \vspace{-1in} \begin{center}
\scalebox{.27} {\includegraphics{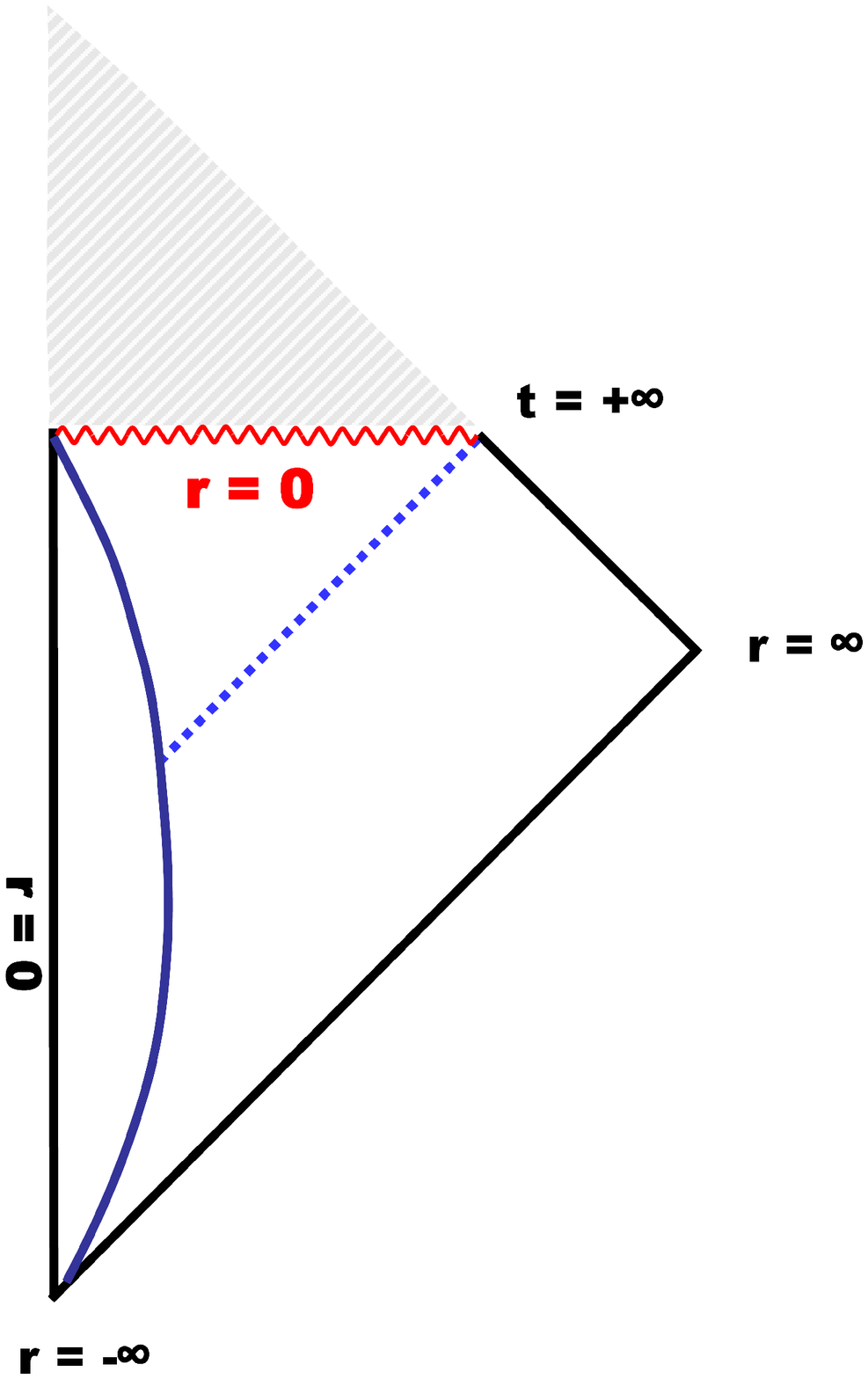}}\ \  \scalebox{.27} {\includegraphics{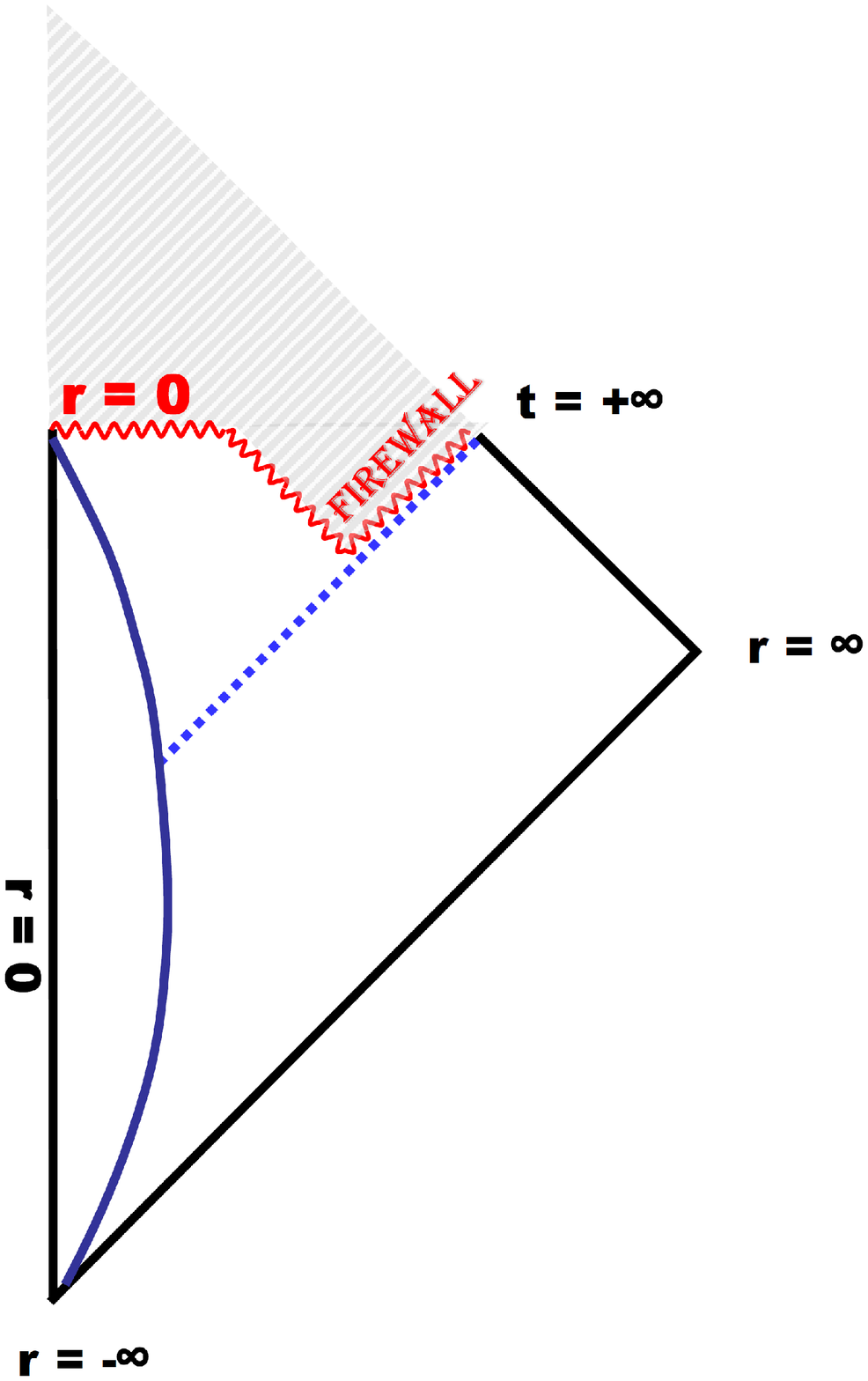}}\\
\scalebox{.30} {\includegraphics{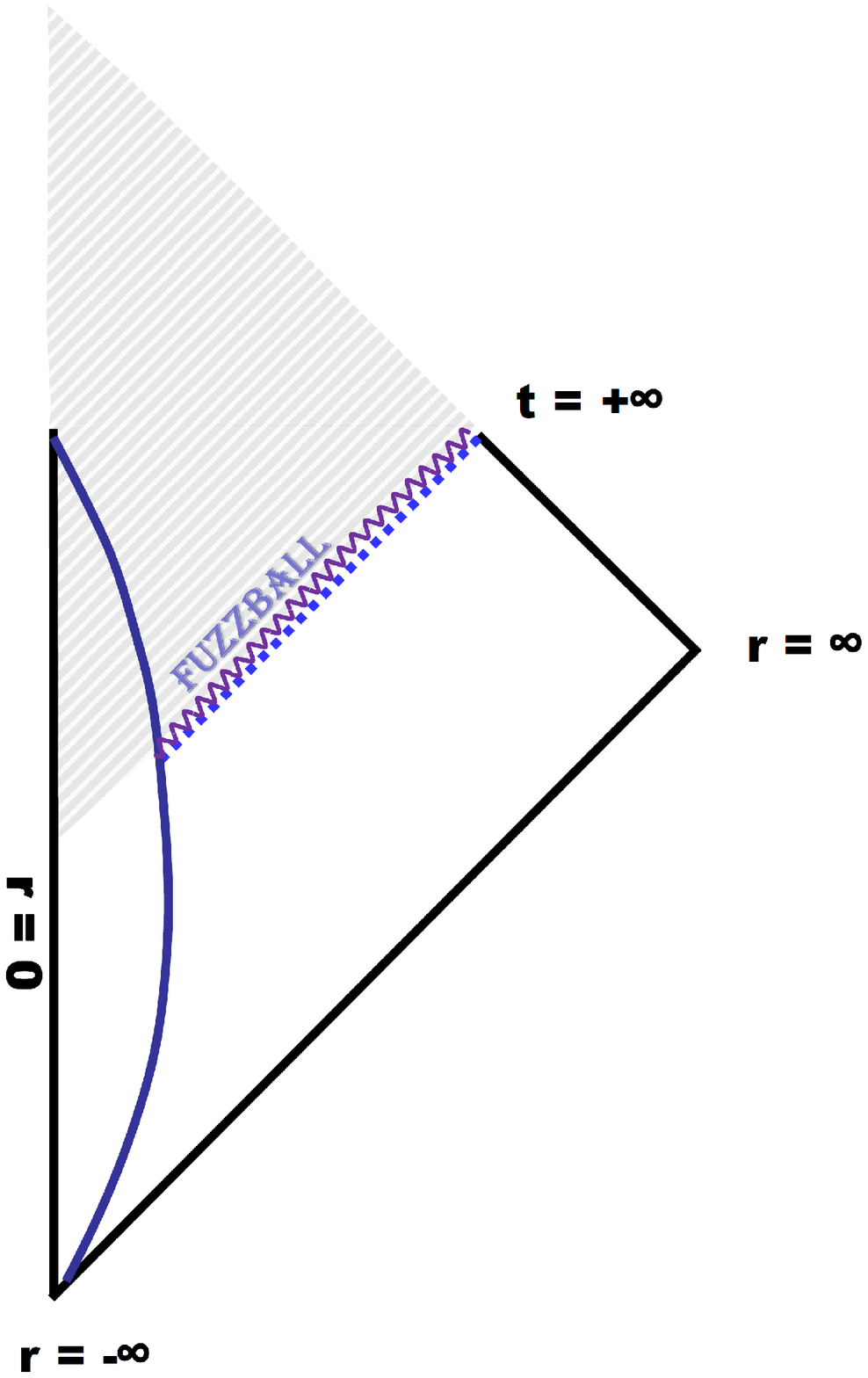}}\ \  \scalebox{.27} {\includegraphics{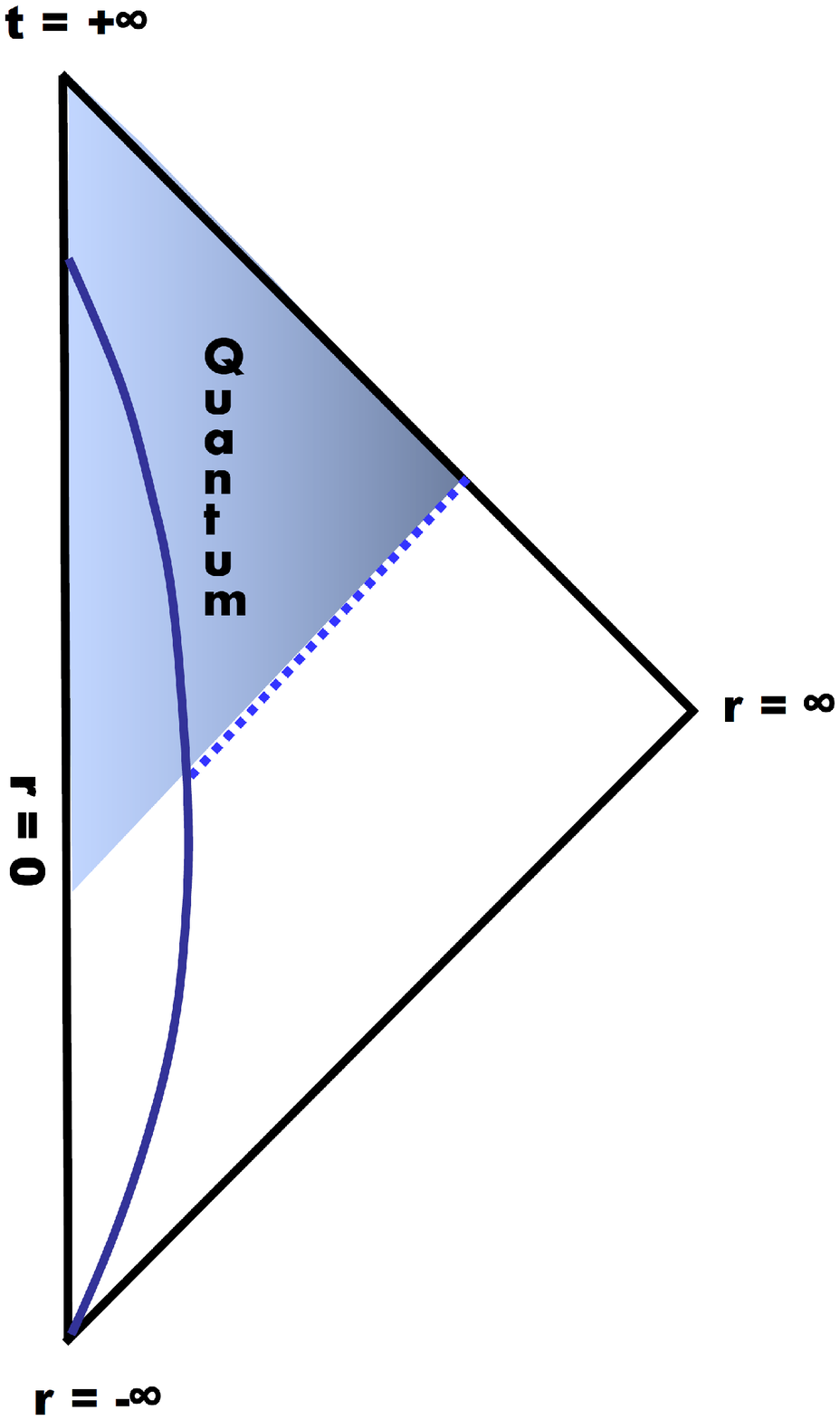}}
\end{center}
\caption{The different options for the BH interior: the classical case (top left), the firewall case (top right), the fuzzball case (bottom left) and the novel quantum case (bottom right).}
\end{figure}

We would like to adopt the viewpoint that the firewall puzzle is a signal that something is amiss with standard semiclassical reasoning and not an indication that the actual physical outcome of the process of BH evaporation is the formation of a firewall at the  horizon. This perspective leads one to a third choice: uniformly distributed matter --- a stance that  has been advocated, in particular, by Dvali and Gomez in  a recent series of articles \cite{Dvali1,Dvali2,Dvali3,Dvali4,Dvali5,Dvali6,Dvali7}. Those authors have reasoned that the BH interior is composed, for the most part, by $\;N\sim S_{BH}\;$ weakly coupled, horizon-sized gravitons  whose state is that of a ``leaky'' Bose--Einstein condensate. Consequently, most of the  gravitons occupy the lowest available quantum state, but their interaction strength is sufficient for gravitons to leak out at a rate that agrees with Hawking's calculation \cite{Hawk, info}.

However, a large-$N$ Bose--Einstein condensate can be effectively described by a coherent state, which has an approximate mean-field description. This is in conflict with the  observation that, whatever composes the BH interior, the state within must be of a highly quantum nature. In fact, Dvali and Gomez observed in \cite{Dvali6,Dvali7} that the condensate description has to break down after the Page time \cite{page}. Our proposal for the state of the BH interior is therefore, in some sense, the ``polar opposite'' of the Bose--Einstein condensate proposal and applies when the condensate picture is no longer valid. However, both proposals share some common scaling relations.

Let us elaborate on the idea of a highly quantum interior state, starting with the observation that, if the emission process of BH radiation is unitary and the initial state is (approximately) pure, then the emitted radiation is itself in a highly quantum state. The BH emissions are sometimes viewed as standard thermal  radiation from a black body on account of the appearance  of  Boltzmann suppression factors, but this picture can be quite  misleading. For instance, the outgoing radiation is practically devoid of particles with non-zero angular momentum, which is much different than what would be expected for, say, particles emerging from a box of radiation.  (The distinction  between BH and thermal radiation is well known and was recently highlighted in \cite{future}.) But the relevant point here is that, since  the BH radiation is in  a highly quantum state, so too must its purifier, the state of  the BH interior. After all,  inasmuch as the initial state of the BH  is (approximately) pure and its evolution is unitary, the reduced density matrices for the radiation and the interior must share the  same set of eigenvalues. This can be viewed as an extension of the ``small-corrections theorem" of Mathur \cite{Mathur1,Mathur2,Mathur3,Mathur4,Mathur5}, which applies to the pair-production picture of BH radiation.

A highly quantum state cannot be described accurately by a mean-field description. In the current context, this means that the state of the matter inside the BH cannot be integrated out and replaced by the gravitational field which it sources. If such a replacement were possible, the inside of the BH could have been described by a semiclassical metric. This is an unusual situation for gravitating systems and, to the best of our knowledge, such states have not been investigated previously.

As  explained later, such highly quantum and highly excited states require a density of states that is much higher than that of normal bound states in known quantum field theories. It is therefore likely that one would need a different framework, such as string theory, for describing these states.

In what follows, we will propose a certain set of scaling relations and then  a matching quantum state for the interior.  This will be followed by a discussion
on various physical aspects of the proposed state, such as its decay via  pair production and the fate of a classical probe that is unfortunate enough to have crossed through the horizon. But, before proceeding, let us comment on the pertinence of our proposal for  the firewall scenario.

The suggestion that a  firewall  is formed around old BHs in \cite{AMPS} and elsewhere is predicated on some common-place assumptions; one of which is the applicability  of standard effective field theory on a fixed and classical curved spacetime with small quantum fluctuations about the background.  This should hold at least as far inward as the so-called stretched  horizon, which is located about a few Planck lengths from the Schwarzschild radius, $R_S$.  It this this assumption that our proposal challenges.  The stretched horizon is already within the region where horizon fluctuations become relevant and, thus, the variance in geometric quantities grows large despite the fact that the curvature and its fluctuations are small. As argued in \cite{RB,flucyou}, this is precisely the point when the above assumption about effective field theory can no longer be trusted. Outside the horizon, on the other hand, we have a good mean-field description of the geometry but {\em not} of the emitted radiation. The Penrose diagram of our proposal in Fig.~1 can be compared to that in \cite{RB}. (See also the similar Penrose diagrams in \cite{Sunny,Vachaspati}.)

Our conventions are the following: Fundamental constants (besides the Planck length, $l_P$) are typically set to unity and we usually neglect  order-unity numerical factors. We consider four-dimensional Schwarzschild BHs for concreteness; however, our results can be extended in a straightforward manner to higher dimensions.

\section{State of the interior}

\subsection{Scaling relations for a maximally entropic state}

Let us consider $N$ relativistic modes whose typical energy $\omega$  is about
$1/R_S$ and for which their total entropy is given by the Bekenstein--Hawking entropy $S_{BH}$,
\be
N \;=\; S_{BH}\;=\; R_S^2/l_P^2
\label{entropy-scaling}
\ee
and their total energy is given by the BH mass $M_{BH}$,
\be
E\;=\;M_{BH}\;=\; N/R_S \;=\; R_S/l_P^2\;.
\label{energy-scaling}
\ee
If these scaling relations hold, then
\be
S(E)\;=\; l_P^2 E^2
\ee
and, as noted in ({\em e.g.}) \cite{MathurX}, the correct thermodynamic relations --- including a negative specific heat ---  will  follow.

We propose a spherically symmetric density of relativistic states that can ``fit" within a given radius and  allow for such scaling relations,
\be
\frac{dN}{dr d\omega}\;=\; \frac{r^2}{l_P^2} \Theta(1-r \omega)\;,
\label{romega}
\ee
from which it follows that
\be
\frac{dN}{dr} \;=\; \frac{r}{l_P^2}
\ee
and
\be
\frac{dN}{d\omega} \;=\; \frac{1}{\omega^3 l_P^2}\;.
\ee
Here, $r$ is the radial coordinate for flat spacetime, which we are using as a fiducial coordinate.

From the above scaling relations, it  can be deduced that, on average,
\be
N(r < R)\;=\; \frac{R^2}{l_P^2}
\label{NlessR}
\ee
and
\be
E(r < R)\;=\; \frac{R^2}{l_P^2} \frac{1}{R_S}\;.
\label{ElessR}
\ee
Meaning that  the energy density and pressure are equal and given by
\be
\rho\;=\; p \;=\; \frac{1}{r^2 l_P^2}\;,
\ee
while the entropy density  is given by
\be
s \;=\; \frac{1}{r l_P^2}\;.
\ee

One can now observe that, in Planck units,
\be
s = \sqrt{\rho}\;,
\ee
which indicates that the causal entropy bound \cite{CEB} is saturated for such a state. Another way to conclude that this entropy bound is saturated is from the equation of state $\;p=\rho\;$. Note that, given our choice for the density of states, the bound is saturated everywhere throughout the interior ---  one cannot find a radius above or below  which  the bound is not  saturated.

In general, the saturation of an entropy bound signifies a breakdown in semiclassical physics~\cite{maxme}. Hence, if the state of the interior satisfies the scaling relations~(\ref{entropy-scaling}) and~(\ref{energy-scaling}), it can not be semiclassical --- it must rather  be a highly quantum state! Consequently, any attempt at  describing it in semiclassical terms is doomed to fail.

The special quantum  nature of this bound state requires a highly dense set of energy levels, as the occupation numbers in each level cannot be large. It follows that the state needs to have at least $S_{BH}$ available energy levels and the spacing between these levels is accordingly very  small,
\be
\delta w \;=\; \frac{1}{R_S}\frac{1}{N} \;=\; \frac{1}{R_S} \frac{1}{S_{BH}}\;.
\ee
In known bound states, the density of states is never this high. Rather, a highly excited state with a large energy and entropy must entail a high occupation  of some of the levels --- the available levels simply run out. It follows that, under normal circumstances, such a state must be semiclassical. However, BHs must be different, as they are highly excited but nevertheless highly quantum.

One can check if such  a state is gravitationally bound  by asking if the total interaction energy between the modes is on par with the total energy of the system. Whereas the latter is $M_{BH}$ by construction, the gravitational  interaction energy of each pair $E^{int}_{ij}$ is approximately equal to
\be
E^{int}_{ij}\;=\; \frac{G_N E_i E_j}{r} \simeq \frac{l_P^2}{R_S^3}\;.
\ee
The total interaction energy is  then given by
\be
E_{int}= \sum_{i,j}^N E^{int}_{ij} \;\simeq\; N^2\times\frac{l_P^2}{R^3_S} \;
\;=\;\frac{N}{R_S}
\;=\; M_{BH}\;.
\ee
From this parametric equality, one can conclude that this is indeed a gravitationally bound state.

One can also ask about the stability of the state, which  can decay through the emission of particles. In principle,  the wavelength of particles $\lambda$ that such a state can emit must be smaller than the Schwarzschild radius, as the  emission of particles with wavelength $\;\lambda>R_S\;$ is highly suppressed due to the geometric suppression factor. In principle, the bound state could  emit particles with wavelength $\;\lambda \ll R_S\;$. However, as we already know and will  discuss later on, the emission of such particles is also highly suppressed, in part, due to the gravitational barrier outside the BH and, in part, because their production rate is suppressed (see Subsection~3.1).  What we then need to know is whether the state is stable against the emission of particles of wavelength $\;\lambda\simeq 1/R_S\;$. Supposing that the state emits a single  particle of energy $1/R_S$, one
  finds that  the scaling relations hold just as well for the resultant state with entropy $N-1$ and energy $M_{BH}-1/R_S$. Then, as the time it takes to emit such a particle is about $R_S$, the state is metastable with a very long lifetime.

\subsection{The quantum state of the interior}

We consider an $N$-particle state of  massless modes (taken
as scalar bosons for  convenience~\footnote{Everything that we say applies just as well
to a state with different species of  bosons and also fermions if light enough species of them exist.\label{foot}}) that are distinguished only by their energies or momenta. The novelty being
that  each energy level
$\;E_i = |\vec{p}_i|\;$
is limited to single occupancy.
Also, the levels are grouped closely around the Hawking temperature $1/R_S$, ensuring that the total energy is on the order of the BH mass as described in Subsection~(2.1).

Formally, this state can be written as
\be
|S\rangle \;\equiv\; \phi^{\dagger}_1\phi_2^{\dagger}\cdots \phi^{\dagger}_{N}|0\rangle\;,
\ee
where $|0\rangle$ is the vacuum state  and each $\phi^{\dagger}_{i}$ is a creation operator with $i$ labeling the $i^{\rm th}$ mode out  of some large number of levels. The label $S$ stands for soliton, the suitability of which will be made clear below.

To  parametrize the desired solitonic behavior,  we can endow each creation operator in $|S\rangle$ with the following Gaussian profile (spherical symmetry is again assumed):
\be
\phi^{\dagger}_i\;= \int\limits^{\infty}_{0} dp\;4\pi p^2 \phi^{\dagger}_i(p)\;,
\ee
where
\be
\phi_i^{\dagger}(p) \;\propto \;e^ {-(p-p_i)^2/(2\widetilde{\sigma}^2_i)}\ e^{i\theta_i}\;,
\ee
with the phase $\theta_i$ allowing for
the possibility of  correlations.

A Fourier transform of this profile leads to
\be
\phi_i^{\dagger}(r)\;\propto\;  e^{-r^2/(2\sigma_i^2)}\ e^{irp_i}\ \phi^{\dagger}_i\;,
\label{mode1}
\ee
where $\sigma_i= 1/\widetilde{\sigma}_i$.
To obtain the scaling relations of the previous section, one needs to choose the momentum-space widths as being on the order of  the Hawking temperature, $\;\widetilde{\sigma}_i\sim \frac{1}{R_S}\;$,
leading to the requisite  position-space width of $\;\sigma_i\sim R_S\;$. Here, instead of the previous $\Theta$-function profile, we have used a Gaussian to obtain a  smoother profile of the state in position space. However, the Gaussian profile should only be regarded as an approximation, as we expect that the true profile is closer to that of the $\Theta$-function of Eq.(\ref{romega}) with some appropriate smoothing near the horizon.

In this way, we arrive   at a picture of $N$ weakly interacting modes having strong support within the horizon
and negligible  support on the outside.  The accompanying integration measure,
\be
\phi_i^{\dagger} = \int\limits^{\infty}_{0} dr\;4\pi r^2 \phi^{\dagger}_i(r)\;,
\label{mode2}
\ee
ensures that the excited modes have roughly uniform support throughout the BH interior, with a peak
probability at a distance $\sigma/\sqrt{2}$ from the origin.

Let us now verify that $|S\rangle$ is a   highly quantum state as prescribed. This can be accomplished by calculating expectation values for field operators such as
\be
\widehat{\Phi}\;=\;\sum_{j}\left[f_j\phi_j +f_j^{\ast}\phi_j^{\dagger}\right]\;,
\ee
where the $f$'s are complex numbers with  order-one magnitude.

Expectation values of odd powers of $\widehat{\Phi}$, including
$\;\langle \widehat{\Phi} \rangle_S\equiv \langle S | \widehat{\Phi} | S\rangle\;$, are all  trivially vanishing and, in a way, this is already enough because
then the ratio of the variance of $\widehat{\Phi}$ to the square of its expectation value
is formally divergent. However, just in case this outcome is non-generic,
let us  consider even powers of $\widehat{\Phi}$.  We have to subtract the vacuum expectation values of the operators, denoted by
$\langle \widehat{O} \rangle_0$, because our interest is only in the contribution from the ``soliton". It is straightforward to show that~\footnote{We are assuming that the operators
for different modes commute and  ignoring
the Gaussian profiles which have no bearing on the current discussion.}
\be
\langle \widehat{\Phi}^2 \rangle_S - \langle \widehat{\Phi}^2 \rangle_0\;\simeq\; 2N\;,
\ee
whereas a longer but still  straightforward calculation leads to
\be
\langle \widehat{\Phi}^4 \rangle_S -\langle \widehat{\Phi}^4 \rangle_0 \;\simeq\; 9 N^2\;.
\ee

One can now see that the variance of the squared operator scales as
(with the vacuum-subtracted definitions now implied)
\be
\frac{\left(\Delta \widehat{\Phi}^2\right)^2_S}{\langle \widehat{\Phi}^2 \rangle^2_S}\;=\;\frac{\langle\widehat{\Phi}^4\rangle_S-\langle\widehat{\Phi}^2\rangle^2_S}{\langle \widehat{\Phi}^2 \rangle^2_S}
\;\simeq\;\frac{5}{4}\;,
\ee
which is, of course,  greater than unity.
The same ratio for higher-order even powers of $\widehat{\Phi}$ will be larger still due to the
growth in combinatorial factors. Simply put, the state $|S\rangle$ is far removed from classical or even semiclassical behavior.

To drive home this last point, we can consider the same basic calculation for the coherent-state analogue of $|S\rangle$,
\be
|\alpha\rangle \;= \;e^{-\frac{1}{2} |\alpha|^2} e^{\alpha\phi^{\dagger}}|0\rangle\;,
\ee
where  $\;|\alpha|\gg 1\;$ to ensure that the number of excitations $N\sim |\alpha|^2$, is large. Note that we have chosen, for simplicity, to excite a single level. Similar results are obtained if many levels are excited, provided that their occupation numbers are large.

Using the defining relation of coherent states, $\;\phi|\alpha\rangle=\alpha|\alpha\rangle\;$, one finds that (with vacuum subtraction again implied)
\be
\langle \widehat{\Phi} \rangle_{\alpha} \;=\; 2\Re({f\alpha})
\ee
and
\be
\langle \widehat{\Phi}^2 \rangle_{\alpha} \;=\;
(2\Re(f\alpha))^2 + {\cal O}(|\alpha|)\;,
\ee
so that the (vacuum-subtracted) variance   scales as
\be
\frac{\left(\Delta \widehat{\Phi}\right)^2_{{\alpha}}}{\langle \widehat{\Phi} \rangle^2_{{\alpha}}}\;\sim\; \frac{1}{|\alpha|^2}\sim \frac{1}{N}\;\ll\; 1\;.
\ee

In any event, if  a quantum state exhibits classical-like behavior,
this ratio is much smaller than unity and, therefore, can function as an order parameter for determining the ``quantumness'' of a  given state. There are times
when higher moments of some operators have small fluctuations; in which case, one can use them as semiclassical order parameters. But, for our solitonic state, no such luck. One simply cannot find a semiclassical order parameter.

It can be  checked that  $\rho$ and $p$ are indeed equal and obey their desired scaling relations,
\be
\rho \;= \; \sum_i \langle \omega_i \phi^\dagger_i \phi_i\rangle_S \;=\; \frac{N}{R_S}\;,
\ee
and similarly for $p$.   The equality
$\;\;p=\rho\;\;$  can also be realized by noticing that
the state has only radial pressure because of its spherical symmetry and that the energy--momentum tensor is traceless because of scale invariance (the modes being relativistic).  So that, effectively, the state is similar to a thermal state in two spacetime dimensions. This analogy will be useful later on.

\section{Physical aspects of the interior state}

\subsection{Pair production}

Let us recall, from \cite{schwing}, a discussion about pair production by a strong gravitational field. In that article,  the gravitational analogue of the  Schwinger effect was considered.  Schwinger's equation \cite{schwinger}  predicts the rate per unit volume $R_{PP}^{\rm E}$ of electron--positron pair production in an electric field ${\cal E}$, where   $\;R^{\rm E}_{PP}= \frac{\alpha^2}{\pi^2 }{\cal E}^2  e^{-\frac{\pi m^2}{e {\cal E}}}\;$. The rate of pair production by a gravitational field is obtained by substituting the gravitational force $F_{\rm G}$ for the electric force $\;F_{\rm E}=e {\cal E}\;$ and the energy $E$ of the positive-energy partner for $m$ in Schwinger's equation.

For our state, the average mass inside a region $r < R$ for $R<R_S$ is found in Eq.~(\ref{ElessR}).
Hence,  the  gravitational force for $R<R_S$ is given by
\be
\;F_{\rm G}(R)\;=\;G_N \frac{E_S(r < R) E}{R^2}\;=\; l_P^2\frac{ R^2}{l_P^2}\frac{1}{R_S} \frac{E}{R^2}\;=\; \frac{E}{R_S}\;,
\ee
and the resulting expression for the rate of  gravitational pair production per unit volume ${\cal R}^{\rm G}_{PP}(R)$ is then
\be
{\cal R}_{PP}^{\rm G}(R)\;\simeq\;  \left(\frac{E}{R_S}\right)^2
e^{-\pi E R_S }\;.
\label{rppg}
\ee

This rate is maximized when $\;E \simeq 1/R_S\;$, which is of the order of the Hawking temperature.  The maximum rate
$\;{\cal R}^{\rm G}_{PP}\simeq {1}/{R_S^4}\;$  agrees with the expectation that one Hawking pair is produced  per light-crossing time $R_S$ from a volume $R_S^3$.

Away from the horizon ($R > R_S$), the rate of pair production becomes exponentially small because the gravitational force becomes small. Inside the horizon, the energy $E$  must be large enough for the positive-energy partner to  ``fit" within this region,  $\;E\;R_S > 1\;$. So that, even in this case, the rate of pair production  is   exponentially suppressed. The latter outcome is contrary to the
very large production rate that   would have been expected if the state had extremely high curvature near the origin.

In summary, a nice consequence of our proposal for the interior state  is that the process of pair production, which  underlies the emission of BH radiation \cite{Hawk,info}, is automatically localized in the vicinity of the horizon.

From this point of view, the net result of (positive-energy) Hawking particles being emitted can be attributed to a standard  application of Fermi's golden rule \cite{Sak}. In particular, the ratio of the rate of emission  of positive-energy modes to their rate of absorption is
\be
\frac{\Gamma_{emi}}{\Gamma_{abs}}\;=\; \frac{\rho_{ext}}{\rho_{int}}\;,
\ee
where $\rho_{ext}$ is the density of states in the exterior spacetime and $\rho_{int}$ is that of the interior region.
Now, as large as the number of level for the interior soliton might be, there is an exponentially  larger number of  available states in the exterior spacetime, as the interior allows only for localized states whereas the exterior encompasses  a continuum (approximately). Hence, the emission of positive-energy modes --- the would-be Hawking particles --- will always be the dominant of the two processes. Put simply, the evaporation of the BH is inevitable.

Finally, let us recall that the stability of the interior state  is maintained as long as all emitted particles carry away a Hawking temperature's worth of energy, as this ensures that the essential relations $\;E\simeq M_{BH}\;$  and $\;N \simeq S_{BH}\;$ are
not jeopardized. In this sense, the whole framework is self-consistent.

\subsection{The plight of an in-falling object}

We next consider  a small classical object, whose linear dimension is $\;L \ll R_S\;$,  falling into the interior bound state.

First, recalling that the energy density of the soliton state is
\be
\rho_S \;\sim\; \frac{1}{ R_S^2 l_P^2}\;,
\ee
we can assign it  an effective temperature of
\be
T_S= (\rho_S)^{1/4} \;\sim\; \frac{1}{ \sqrt{R_S l_P}}\;.
\label{temp}
\ee
Here, we are ignoring the fact that the equation of state is $\;p=\rho\;$, rather than $\;p=\frac{1}{3} \rho\;$, because the detailed interaction between the soliton and the object is not important.

What is important is that the fluctuations in energy density and pressure are relatively large (see Subsection 2.2) and that the various light particles in the interior region (photons, gravitons, {\em etc.}) participate democratically
in the ``thermalizing'' process. Hence, if the falling object  is ``colder" than the effective temperature, it will heat up at a rate which is approximately given by
\be
\frac{d E}{dt} \;\simeq\; T_S^4 L^2 \;.
\ee

To get an idea about the scale of the temperature and energy absorption, let us  consider a solar-mass BH. Then the effective temperature can be parametrized as
\be
T_S\;\simeq\; 1\; {\rm GeV} \sqrt{\frac{R_S}{(R_S)_{\bigodot} }}\;,
\ee
where $\;(R_S)_{\bigodot}\simeq 3\times 10^3$~m is the Schwarzschild radius of a
solar-mass BH.
It is clear that any classical object falling into this state will heat up quickly and disintegrate. To have an effective temperature as ``low'' as $\;1000^\circ$~K $\sim 0.1$~eV, the BH radius would have to be  larger than $(R_S)_{\bigodot}$
by a factor $10^{20}$; that is,  almost as large as the whole universe!  Of course, this is an extremely rough estimate, but it nevertheless shows that any classical object will not survive the extreme atmosphere of the BH interior.

\subsection{Mean free path of a spherical wave}

In general, the mean free path (MFP) of a wave in a random medium can be defined as the length  $\ell$ that is traversed by the wave when  the expectation value for  a collision with the medium becomes of order one,
$
g n \lambda^2 \ell \;=\; 1\;.
$
Here,  $g$ is the dimensionless coupling between the wave and the medium, and
can be thought of as the probability for the wave to interact with a scatterer in the medium, $n$ is the number density of scatterers  and
$\lambda^2\;\ell$ is the volume that the wave sweeps out
 ---  the cross section of the wave times the length that it traverses between interactions.

In our case, we consider a spherically symmetric distribution of scatterers ---
 the modes of the interior whose number density is given by Eq.~(\ref{romega}). Therefore, the appropriate definition of the MFP is given by the following condition:
\be
\int_{r_{initial}}^{r_{initial}+\ell} dr d\omega\ g(r,\omega)\; \frac{d^2 N}{dr d\omega} \;=\;1\;.
\label{defmfp}
\ee

Let us consider a spherical wave  propagating outwards through the BH interior at the speed of light with energy $E_{wave}$. We will assume that the wave interacts only gravitationally with the medium. Of course, if it also interacts  electromagnetically or via the strong or weak forces, then the MFP would be much shorter than the following estimate.

The wave must be initially of size $\;R <R_S\;$ with an  energy $E_{wave}$ satisfying  $\;E_{wave} R_S > 1 \;$.
The coupling of the wave with the medium is given by
\be
g(r,\omega)\;=\; l_P^2 E_{wave}\ \omega \;,
\ee
so that Eq.~(\ref{defmfp}) becomes
\be
\int_{r_{initial}}^{r_{initial}+\ell} dr d\omega\ l_P^2 E_{wave}\ \omega  \frac{r^2}{l_P^2} \Theta(1-r\omega) \;=\;1\;.
\label{mfpwave}
\ee
The powers of $l_P$ cancel in Eq.~(\ref{mfpwave}), and so the MFP
can depend only on $E_{wave}$ and  $R_S$.

Performing the $\omega$-integral, we obtain
\be
\int_{r_{initial}}^{r_{initial}+\ell} dr E_{wave} \;=\;1\;,
\label{mfpwave1}
\ee
from which it follows that
\be
\ell\; =\; \frac{1}{E_{wave}}\;.
\label{mfpresult}
\ee
An immediate implication is that, if $\;E_{wave} R_S \gg 1\;$,
 then $\;\ell \ll R_S\;$.

In the limiting case, when the energy of the wave is of order $\;E_{wave}\sim 1/R_S\;$, then $\;\ell \sim R_S\;$.
The conclusion here is  that any spherical wave which  is  small enough to  fit
within the interior region will be stopped at a  distance that is
parametrically  smaller than $R_S$, even if {\em only} gravitational interactions  are considered.  Thus, the only viable means for escape from the interior is through the process of pair production
at the horizon; that is, through  the ``standard''  process of Hawking radiation.

As mentioned above,  by neglecting the non-gravitational forces, we have greatly underestimated the stopping power of the interior matter. The point is that the number density in the interior $n$ scales with $l_P^{-2}$ whereas a typical force (like the electromagnetic force) has an interaction strength of order unity. Consequently, the MFP can roughly be expected to go as
\be
\ell\;\lesssim\; \frac{1}{f} R_S \; l_P^2 E_{wave}^2 \;,
\ee
where $f$ indicates the order-unity fraction of the interior constituents that participate in non-gravitational interactions. It is straightforward to confirm that  $\ell\ll R_S$ for any realistic choices of $f$ and $E_{wave}$.

\section{The end of semiclassical space}

Our expectation is that the semiclassical approximation is no longer
trustworthy already at    a short distance away  from the classical location of the horizon ($r = R_S$). What we would then like to know is just  how far past the horizon does the semiclassical approximation remain (nearly) valid.
That is, where in the interior does the picture of semiclassical physics break down completely?  Unfortunately, we cannot answer this question with a calculation but find that it can still be addressed on a qualitative level. For this, we
will  appeal to the ER=EPR perspective of entanglement \cite{ER}; namely, that the amount and complexity of the pattern of entanglement between the BH and the outgoing radiation is somehow connected to the nature of the interior geometry.

Let us define $\Delta R_{SC}$ as the distance scale that determines  how far past the horizon
one encounters the end of semiclassical spacetime. (See Fig.~2.) Our  premise is then  that this distance scale is determined by the amount of disentanglement for the produced Hawking pairs near the horizon. In \cite{stick}, we have defined this amount by $N_{dis}$, whose  value depends on the particular model of BH evaporation.
So that, here, large $N_{dis}$ is associated with small $\Delta R_{SC}$ and {\em vice versa}.

We will consider three different cases: \\

\noindent
(1) If $\;N_{dis} \sim 1\;$, then $\;\Delta R_{SC} \sim R_S\;$ and the state of the emitted radiation and the state of the interior are both mixed. In this case, all of the interior is semiclassical up to (about) the singularity. With this choice of $N_{dis}$, we
are essentially describing the Hawking model of BH evaporation \cite{Hawk,info},
as the produced  pairs are presumed to be maximally entangled in this case. \\

\noindent
(2)
If $\;N_{dis} \sim S_{BH}\;$, then $\;\Delta R_{SC}\sim l_P\;$ and the purifier of the radiation is the ``energetic curtain" near the horizon of the BH. Then
only  a Planck-sized layer near the Schwarzschild radius is semiclassical. With this choice of $N_{dis}$, we are essentially describing
the Page model of BH evaporation \cite{page} or, more accurately, how this model has been
interpreted in the firewall literature. It is then not a coincident that
the resulting picture looks very much like a firewall close to the horizon, and one should think about the interior as being disconnected from the outside world. \\

\noindent
(3)
If $\;N_{dis} \sim \sqrt{S_{BH}}\;$,  then we expect that $\Delta R_{SC} \gg l_p$ but also that $\Delta R_{SC} \ll R_S$. A possible choice could be the ``thermal wavelength'' of the interior state  $\;\Delta R_{SC} \sim \sqrt{R_S l_P}\;$ ({\em cf}, Eq.~(\ref{temp})). Clearly, for distances that are parametrically larger than this wavelength, the interior state cannot be described in semiclassical terms. However, it may well be that, for some reason, $\Delta R_{SC} \ll \sqrt{R_S l_P}$ or that the profile of the interior state in the vicinity of the horizon is not generic.

\begin{figure}
[t]
\begin{center}\scalebox{.45} {\includegraphics{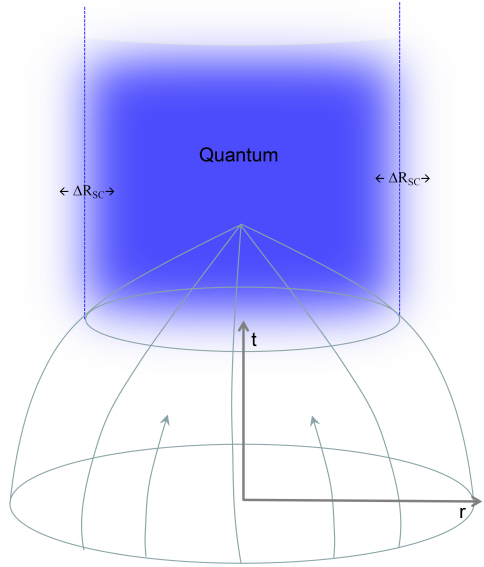}}\end{center}
\caption{Spacetime diagram of a collapsing shell and the resulting BH.  The end of semiclassical spacetime is encountered at a short distance
$\Delta R_{SC}$ away from the BH horizon.}
\end{figure}

If indeed $\;\Delta R_{SC} \sim \sqrt{R_S l_P}\;$, then the interior still looks  semiclassical for a distance away from the horizon that is  parametrically small  compared to $R_S$ but parametrically large  compared to the Planck length.   To make this estimate a little bit more quantitative, we can parametrize it as  $\;\Delta R_{SC} \sim 0.1\;{\rm fm} \left(R_S/(R_S)_{\bigodot}\right)^{1/2}\;$; so that, even in the case of a macroscopic BH, the distance is subatomic!  This choice of $N_{dis}$ is consistent with a recently proposed  semiclassical model of  BH evaporation  \cite{slowleak,slowburn,flameoff}. \\

The reader can consult \cite{future} for  further  discussion that
compares and contrasts the three cited models. It should be noticed, however,
that only the semiclassical model of Case~(3) is consistent with the
highly quantum interior state that has been proposed in this paper.
The inapplicability of  Case~(1) should be obvious, while
that  of  Case~(2) is almost as simple: If there are already order-$S_{BH}$
disentangled Hawking pairs in the vicinity of the horizon, then there is no longer the possibility
of having $S_{BH}$ modes of total energy $M_{BH}$  within it.

\section{Conclusion}

The main idea that is put forward  in this paper is that the BH interior is in a highly quantum state. From this point of view, the long debate about the information paradox was misplaced. Rather than questioning whether or not the evolution is unitary, one should concentrate on finding the quantum state of the BH in the vicinity of the horizon and  its interior. There is no {\em a priori} reason that the interior  state has to  be semiclassical. As the detailed nature of this state is still not clear,  our proposal should be viewed as an initial, perhaps crude attempt at modeling what the BH interior could resemble. Nevertheless, our model of a ``Fermi sea'' of light-enough particles is appropriately quantum, provides an accounting of the BH mass and entropy and even allows for pair production to occur at the horizon in a natural way.

Normal bound states in known quantum field theories do not seem to be able to reproduce the high density of states which is required to describe the interior of BHs. It would therefore be interesting to determine whether string theory allows for such  highly excited and highly quantum bound states with the required density of states.

It is a challenge to come up with direct observable consequences for our model, inasmuch as the mean-field approximation fails   for such  highly quantum states. In particular, it would be interesting to think about some tests that will be able to differentiate our proposal from the fuzzball or the firewall proposals. Theoretically, these proposals are quite different from ours in terms of the structure of the near-horizon region. For instance,  from our perspective, the safety  of the classical probe is compromised {\em but} only after falling through a distance which is parametrically larger than Planckian past  the horizon and  into a region where the semiclassical
approximation is no longer applicable. However, in practice, the differences are more subtle and further attention is warranted.

\section*{Acknowledgments}

We would like to thank Sunny Itzhaki for valueable discussions and Gia Dvali and Samir Mathur for useful comments on the manuscript. We are indebted to Gabriele Veneziano
for discussions and for sharing his preliminary results in a parallel
investigation. The research of RB was supported by the Israel Science Foundation grant no. 239/10. The research of AJMM received support from an NRF Incentive Funding Grant 85353, an NRF Competitive Programme Grant 93595 and Rhodes Research Discretionary Grants. AJMM thanks Ben Gurion University for their  hospitality during his visit.

\end{document}